# Two-photon Lithography for 3D Magnetic Nanostructure Fabrication

Gwilym Williams[1], Matthew Hunt[1], Benedikt Boehm[2], Andrew May[1], Michael Taverne[3], Daniel Ho[3], Sean Giblin[1], Dan Read[1], John Rarity[3] , Rolf Allenspach[2] and Sam Ladak

[1] School of Physics and Astronomy, Cardiff University, Cardiff U.K.
[2] IBM Research - Zurich, Säumerstrasse 4, 8803 Rüschlikon, Switzerland
[3] Department of Electrical and Electronic Engineering, University of Bristol, Bristol U.K.

# Two-photon Lithography for 3D Magnetic Nanostructure Fabrication

Gwilym Williams[1], Matthew Hunt[1], Benedikt Boehm[2], Andrew May[1], Michael Taverne[3], Daniel Ho[3], Sean Giblin[1], Dan Read[1], John Rarity[3], Rolf Allenspach[2] and Sam Ladak[1] (Email: LadakS@cardiff.ac.uk)

[1] School of Physics and Astronomy, Cardiff University, Cardiff U.K.
[2] IBM Research - Zurich, Säumerstrasse 4, 8803 Rüschlikon, Switzerland
[3] Department of Electrical and Electronic Engineering, University of Bristol, Bristol U.K.

## ABSTRACT

Ferromagnetic materials have been utilised as recording media within data storage devices for many decades. Confinement of the material to a two-dimensional plane is a significant bottleneck in achieving ultra-high recording densities and this has led to the proposition of three-dimensional (3D) racetrack memories that utilise domain wall propagation along nanowires. However, the fabrication of 3D magnetic nanostructures of complex geometry is highly challenging and not easily achievable with standard lithography techniques. Here, by using a combination of two-photon lithography and electrochemical deposition, we show a new approach to construct 3D magnetic nanostructures of complex geometry. The magnetic properties are found to be intimately related to the 3D geometry of the structure and magnetic imaging experiments provide evidence of domain wall pinning at a 3D nanostructured junction.

## 1. Introduction

Current information technologies are reliant upon growth and processing techniques that allow the fabrication of two-dimensional nanostructures [1]. This currently involves the exposure of a photoresist to light through a mask to generate a nanoscale pattern, which is subsequently transferred into another material via deposition or etching processes [1].

The increasing demand for higher data storage densities whilst also maintaining short access times has recently generated an interest in three-dimensional (3D) data storage solutions such as magnetic racetrack memory [2]. On the other hand, the experimental and theoretical study of 3D magnetic nanostructures is also of fundamental interest and has recently allowed the probing of new and exciting physics such as Bloch point domain wall propagation [3], curvature induced effective Dzyaloshinkii-Moriya interaction [4] and magnetic charge transport in 3D artificial spin-ice structures [5]. These studies have required alternative fabrication strategies in order to realise very simple 3D geometries. A common approach to fabricate 3D magnetic nanostructures is to electrodeposit magnetic material into alumina templates or ion-track templates. This is a promising route since electrodeposition is a well-established method of growing a range of magnetic materials [6] and has allowed the fabrication and characterisation of cylindrical magnetic nanowires [7] [3]. More recently ion track etched templates have also been used to fabricate interconnected networks of NiCo nanowires [8] but unfortunately the technique does not currently allow the fabrication of complex 3D geometries by design. Focused electron beam deposition is a powerful technique that has demonstrated the fabrication of 3D magnetic nanostructures and has yielded geometries such as wires and helices [9]. The method can be used to fabricate a range of magnetic materials [9] and has also recently led to structures with measurable magneto-optical Kerr effect signals [10]. However, the deposited material has often been contaminated with large amounts (>5%) of carbon and oxygen. It is also difficult to envisage how the technique would be able to fabricate complex, extended 3D networks or geometries. In other disciplines, a range of self-assembly and chemical methods have yielded great success in the fabrication of 3D nano/microstructures including chiral liquid crystal structures [11,12] and hybrid 3D graphene gold nanoparticle structures [13]. In addition, the manipulation of droplets upon surfaces via magnetic guiding and three phase contact lines has been shown to be a powerful means to produce 3D microstructures in a number of geometries from magnetic ink [14], CdTe quantum dots [15], silver nanoparticles [15] and manganese chloride salt [15].

Two-photon lithography (TPL) [16] is a relatively new technique that has largely been exploited within the metamaterials and microfluidic communities to fabricate complex 3D nanostructured materials. In this technique a femtosecond laser operating in the infra-red (typically $\lambda \approx 780nm$) is focused to a diffraction-limited spot within a conventional photoresist. The high peak intensity at the focal point allows the simultaneous absorption of two-photons in order to excite the electronic transition within the photoinitiator molecule, leading to polymerisation or depolymerisation of the resist. This non-linear optical process is proportional to the intensity squared and thus only occurs within the central region of the focal spot. By moving the point of focus within the resist, 3D nanostructures of arbitrary geometry can be produced. The technique can be used to manufacture 3D nanostructures within a polymer or by employing surfactant-assisted multiphoton-induced reduction, the realization of metallic nanostructures [17].

In this study we utilise a novel fabrication approach using a combination of TPL and electrodeposition to fabricate complex 3D Co magnetic nanostructures by design. The structures are of high purity and can be measured using standard surface sensitive techniques such as the magneto-optical Kerr effect (MOKE) and spin-polarised scanning electron microscopy (spin-SEM). Our technique provides a new route to fabricate 3D nanomagnetic elements and wires with bespoke properties.

## 2. Results and Discussion

An overview of the fabrication procedure is shown in Figure 1(a-d). A positive photoresist is spun onto a glass/ITO (500nm) substrate (Fig 1a) and TPL is used to expose a 3D pattern within the resist (Fig 1b). Development removes the exposed regions of the resist, leaving a series of channels. Electrodeposition is then used to fill the channels with Co (Fig 1c) after which the resist is removed yielding a 3D magnetic nanostructure (Fig 1d). In order to investigate the possible lateral feature size, arrays of cylinders were fabricated where the power and development time were varied. The challenge in fabricating complex 3D geometries within a positive resist is balancing a development time that allows removal of exposed resist, whilst also minimising dark erosion which leads to larger feature sizes. Fig 1(e) shows the variation of lateral feature size with laser power for a fixed development time of 30 minutes. The lateral feature size decreases non-linearly with laser power and the minimum lateral feature size is found to be approximately 435nm.

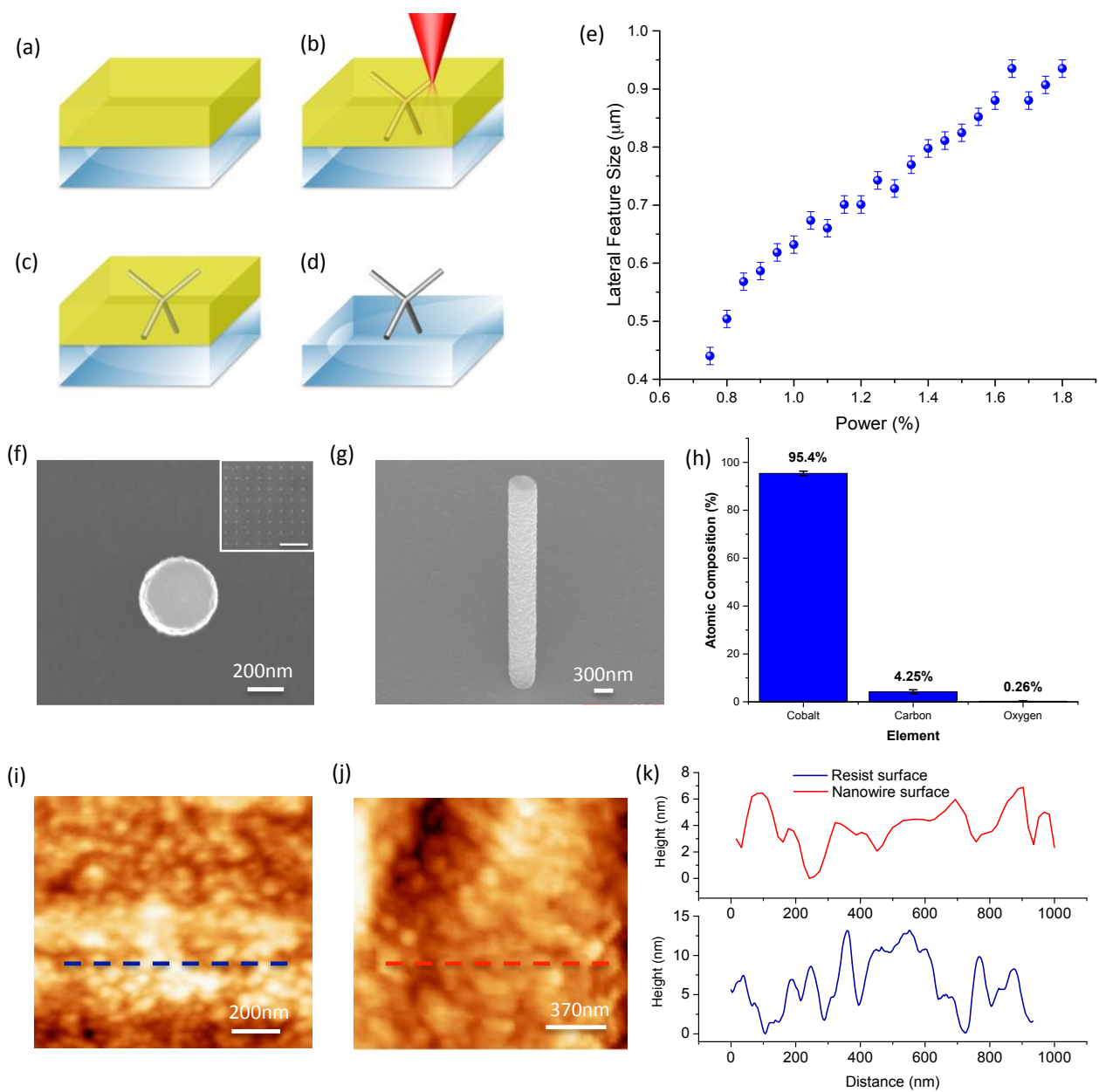

**Figure 1 Fabrication and structural characterization of 3D magnetic nanostructures.** (a) Spinning of positive resist onto glass/ITO substrate. (b) Two-photon lithography of 3D structure into positive resist. (c) Electrodeposition of Co into the channels. (d) Lift off of resist. (e) The lateral feature size obtained for pillars fabricated at different powers. (f) SEM micrograph of a single 435 nm Co nanowire. Inset: Array of sub-500 nm pillars. Scale bar represents 8 µm (g) SEM micrograph of a single 435 nm Co nanowire taken at 60-degree angle. (h) Energy dispersive X-ray analysis of Co pillars. (i) Atomic force microscopy image of resist channel surface. (j) Atomic force microscopy image of a nanowire sidewall. (k) Profiles taken from nanowire sidewall and resist channel atomic force microscopy images (i,j).

Simple models of the two-photon lithography process within a positive resist show that the channel diameter D can be expressed by [18]:

$$D = w(z)\sqrt{\ln\left[\frac{4C\eta^2 P_{laser}^2 t}{f\tau(\pi h\nu w^2(z))^2 \ln\frac{M_0}{M_{th}}}\right]} \quad (1)$$

where z is the distance from laser focus, w(z) is the beam radius at position z, C is a factor relating to the product generation rate, η is the transmittance of the objective lens, $P_{laser}$ is the incident laser power, t is the processing time, f is the repetition frequency of the laser, τ is the pulse width, ν is the frequency of light, $M_0$ is the initial concentration of photoinitiator in the ground state and $M_{th}$ is the threshold amount of dissolvable photoinitiator. With our optical parameters (see methods) this yields a minimum feature size of approximately 280nm. However, taking into account a dark erosion rate of ≈5nm per minute [19] yields a feature size of approximately 430nm, close to the observed value.

Fig 1(f) and Fig 1(g) show a SEM image of a 435nm diameter Co nanowire. The nanowire is well defined with circular cross-section, smooth top and has length 3µm. The inset of Fig 1(f) shows that it is straightforward to fabricate large, regular arrays of sub-500nm nanowires. Energy Dispersive X-ray analysis was performed upon larger 1 µm structures (in order to minimise the signal from the ITO substrate). Figure 1(h) shows the elemental composition obtained after averaging over ten structures and subtracting a small background from the substrate. The structures are found to be of high purity with a Co composition of >95%. The small amount of carbon and oxygen detected is due to the SEM process and a small amount of residual resist.

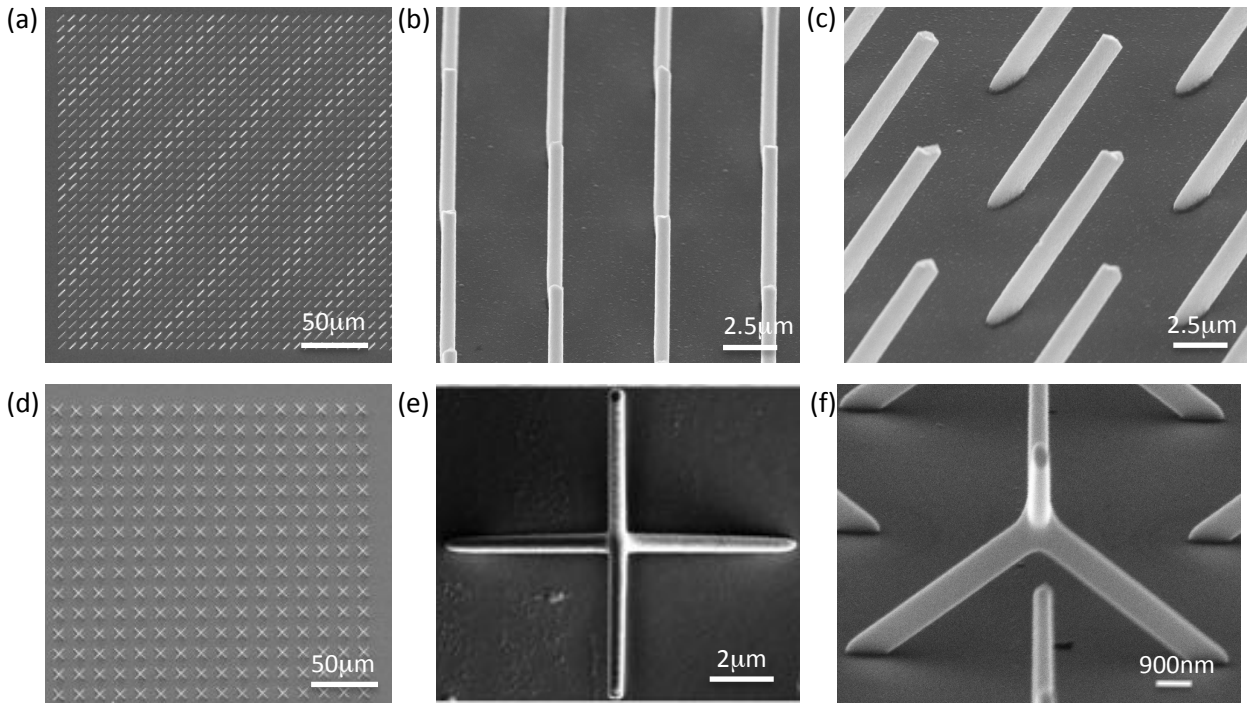

**Figure 2**. **Scanning electron microscopy of 3D magnetic nanostructures** (a) SEM image of a tilted nanowire array (Top view). (b) High magnification image of tilted nanowire array. (c) SEM image of tilted nanowire array taken after 90° in-plane rotation. (d) Large scale SEM of tetrapod array (Top view). (e) High magnification image of single tetrapod. (f) SEM image of tetrapod taken after 45° out-of-plane substrate rotation.

Surface roughness is an important factor in determining the final properties within magnetic nanostructures. Within our structures there are three surfaces that need to be considered. The surface making contact with the substrate is likely to be limited by the roughness of the underlying surface as in standard thin films. However, the surface roughness of channel sidewalls and their impact upon the nanowire morphology has not been studied previously. In order to investigate this, in-plane channels within the resist were fabricated using two-photon lithography and were measured using atomic force microscopy (AFM) as shown in Fig 1i. This was compared with the surface roughness of a microwire (diameter≈1.5μm) that had fallen, exposing its sidewalls (Fig 1j). We find that the surface roughness of the microwire (3nm) is very close to that of the channel surface (5nm).

Hence it is likely that the morphology of the channel sidewalls constrain the edge roughness of grown nanowires. Finally, the upper surface of the nanowire is likely to be a strongly dependent upon the electric field line distribution across the channel. Previous studies have already demonstrated, via an active-area density model [20], that current crowding effects can lead to non-uniformities in the upper surface of electrodeposited structures. The effect is most pronounced within larger microstructures where current crowding at the channel edges leads to a thicker region at the electrolyte-resist interface. This is less pronounced in our smallest (430nm diameter) structures but is mildly apparent in medium-sized (600-700nm) structures and more pronounced in micro-metre sized structures.

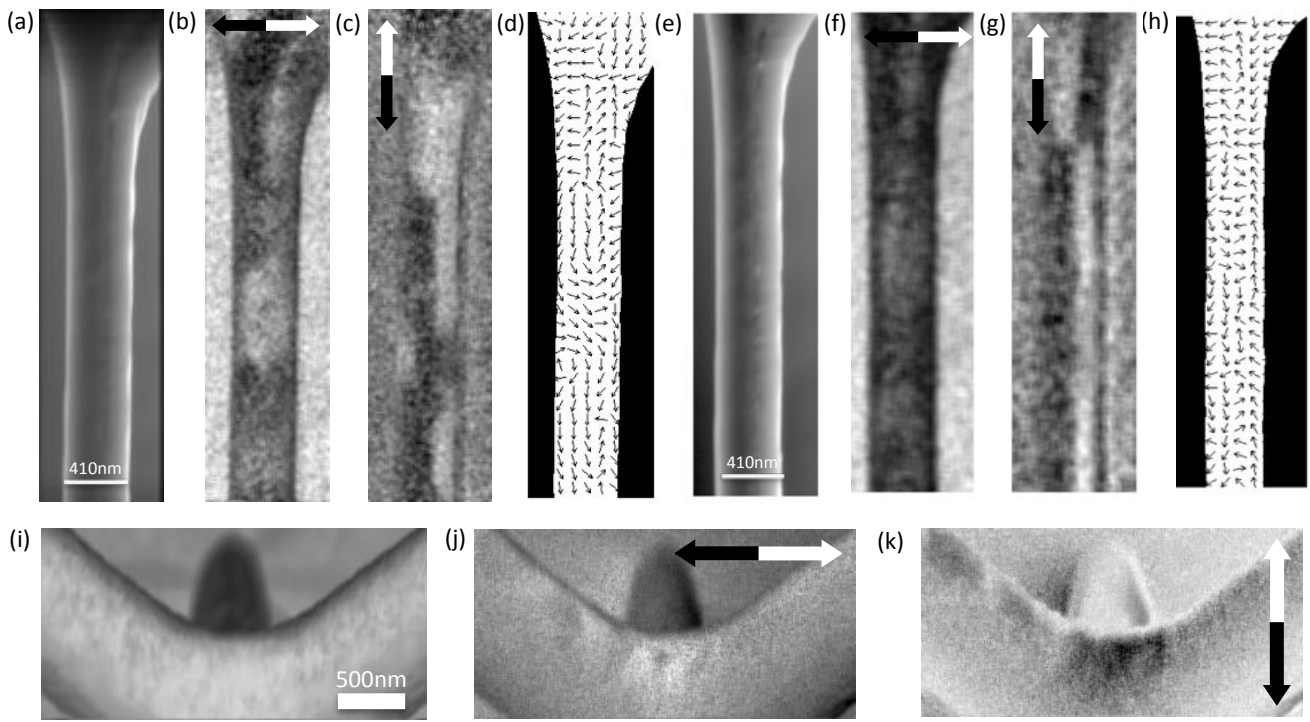

**Figure 3: Spin-SEM micrographs of 3D magnetic nanostructures** (a) Absorbed current image of individual wire within tetrapod structure. Spin-polarised SEM image showing (b) x-component and (c) y-component of magnetisation in as deposited sample; (d) direction of in-wire magnetisation as deduced from (b) and (c) for the as-deposited sample. (e)-(h) Same sequence as (a)-(d) after magnetic flux density of 11.8 mT. Spin-SEM micrographs of vertex area showing magnetisation contrast centred upon the vertex area; (i) topography, (j) x-component and (k) y-component of magnetisation.

In order to demonstrate the versatility of our technique in fabricating 3D nanomagnetic structures, we have used TPL in order to fabricate arrays of angled wires and then used this design as a building block to realise complex 3D tetrapod structures. The approach is particularly powerful since it should help to understand the switching of a complex 3D magnetic nanostructure in terms of its underlying constituents. Fig 2(a) shows a ~ 300 μm × 300 μm array of angled wires. The array is well ordered with no defects and low distribution of wire lengths. High magnification images of the wires are shown in Fig 2(b) and Fig 2(c). The wires have an elliptical cross section with semi-axes of 660nm in the substrate plane, 885 nm perpendicular to the wire long axis, a length of ≈8 μm and make an angle of 30.5° with respect to the substrate. The elliptical cross section of the wire is due to the geometry of the point spread function at the focal point of the objective during TPL [21].

Fig 2(d) shows a SEM of a 300 μm × 300 μm array of 3D tetrapod structures that were fabricated with parameters similar to the single wires. The high magnification SEM images shown in Fig 2(e) and Fig 2(f) demonstrate that the complex tetrapod geometry has been successfully realised. The wires within the tetrapod structures are found to have a length of approximately 8 μm and an elliptical cross section with semi-axes of 615 nm in substrate plane and 853 nm perpendicular to the wire long axis. Before attempting surface magnetometry upon the 3D nanostructured samples it is useful to have some understanding of the underlying domain structure upon the tetrapod surface. Most standard magnetic imaging techniques such as magnetic force microscopy (MFM) [22], photoemission electron microscopy (PEEM) [23] and Lorentz microscopy [24] are not well suited to studying 3D nanostructured samples. Here we exploit the large depth of focus in spin-SEM in order to image a 3D nanostructured magnetic sample. Figs 3(a), 3(b) and 3(c) show the topography and the x- and y-components of magnetisation respectively, taken upon a single wire within an as-deposited tetrapod structure using spin-SEM. An arrow representation of the x- and y-components is displayed in Fig 3(d). Figs 3(e)-(h) show the same wire after application of an in-plane pulse with magnetic flux density of 11.8 mT along the projection of the long axis upon the substrate. The images clearly show that the system is multi-domain both before and after application of a field, with the remanent magnetisation being larger along the wire long axis. The demagnetisation field perpendicular to the long axis within a tetrapod nanowire is expected to be strong ($0.98 M_s \cdot \mu_0 = 1.7$ T) and this will lead to the magnetisation mainly lying parallel to the long axis at remanence, as seen in Fig 3(g). However, an appreciable angular spread of magnetisation direction is observed in Fig. 3(h). Electrodeposited hcp Co is also expected to have an uniaxial magnetocrystalline anisotropy directed along its c-axis with $K_1 = 5.3 \times 10^5$ J/m$^3$. This is approximately a factor of 4 lower than the energy associated with the demagnetisation field ($2.3 \times 10^6$ J/m$^3$). Hence, for randomly oriented crystallites within our wires, we expect the anisotropy term to lead to a distribution of magnetisation angles with respect to the long axis.

Magneto-optical Kerr effect (MOKE) magnetometry has been carried out in the longitudinal geometry upon both samples sets [25]. Fig 4(a) shows a hysteresis loop that was measured with the field along the projection of the wire long-axis onto the substrate surface and with angle of incidence of 14.5° with respect to the wire long axis. The loop displays a steep rise at low fields, which is followed by a slow gradual approach to saturation. The remanence is rather small, around $0.3 M_s$, in line with published values [26]. Fig 4(b) shows a hysteresis loop that was measured with the field perpendicular to the projection of the wire long-axis onto the substrate. The curve resembles more a hard axis magnetisation curve without reaching saturation at 0.5 T, the largest field available in our setup. We infer that the curve is dominated by rotation of the magnetisation, required to overcome the strong demagnetisation field perpendicular to the wire long-axis, typical of electrodeposited Co nanowire hard-axis loops [27]. It has been reported that electrodeposited nanowires have crystalline grains with a spread in c-axis direction, but there is no general agreement on the preferential orientation of the c-axis [27] [26]. However, some studies upon electrodeposited Co within sulfate-based baths have demonstrated pH can be an effective means to tune the crystallographic orientation and hence magnetocrystalline anisotropy [27]. Fig 4(c) shows the longitudinal MOKE loop that is obtained upon the tetrapod array when the magnetic field is parallel to the projection of the lower wire long-axes. At low fields, the magnetisation displays a much more abrupt switching than the single-wire array shown in Fig 4(a). Further smeared transitions at higher fields can also be discerned. A rather similar loop was obtained when the magnetic field was applied perpendicular to the projection of the lower wires, see Fig 4(d). The coercivity in this case is seen to be lower than in Fig 4(c) but again smaller transitions can be seen at fields above the coercive field.

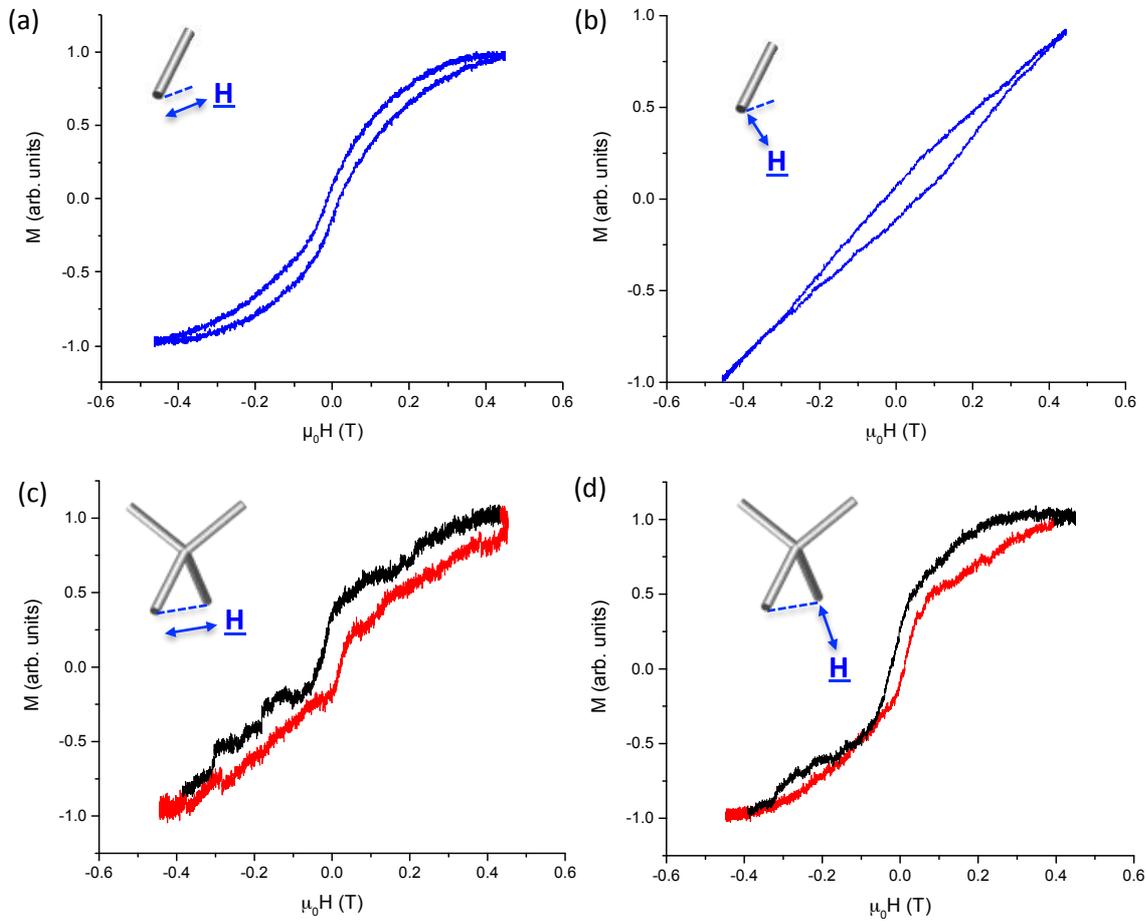

**Figure 4: Magneto-optical Kerr effect magnetometry.** Longitudinal MOKE loops taken upon (a) single wire array with field applied along projection of wire long axis, (b) single wire array with field applied perpendicular to projection of wire long axis, (c) tetrapod array with field applied along projection of lower wires and (d) tetrapod array with field applied along projection of upper wires.

One can gain some insight into the reversal mechanism within tetrapods by considering the planar equivalent (sub-micron crosses) of our structures that have been studied previously [28]. In these samples the nanostructuring was found to play an important role and a complex set of 180° and 90° domain walls was found to form at the vertex area, during the magnetic reversal process. In our tetrapod geometries a four-way junction is also present so it seems likely that domain wall pinning at the vertex area will impact the observed magnetometry and may be responsible for multiple switching events. For the tetrapod loop shown in Fig 4(c), where the field is parallel to the projection of lower wires, an initial transition ($H_{c1}$) occurs at a field of 21mT with only a small variation seen in nominally identical experiments (±2mT). This is close to the coercive field observed for single wires (Fig 4a) with field parallel to projection of long axis (17±2mT). It therefore seems likely that the initial transition seen in Fig 4(c) is due to switching of lower wires that have component parallel to field, after which the 3D nanostructured vertex impedes domain wall movement into the upper wires. Further transitions are consistently seen above $H_{c1}$. Here, it is not clear if a domain wall depinning event occurs at the vertex, or if magnetisation rotation in the upper wires start to become significant enough to be observed in the magnetometry. Stochasticity of domain wall processes is likely to yield a distribution of depinning fields above $H_{c1}$, as observed in Fig 4c.

It is likely that larger structures of similar geometry will not be susceptible to geometric domain wall pinning processes due to nanostructuring. In order to test this hypothesis we fabricated much larger tetrapod structures that have a wire length of approximately 6 μm and an elliptical cross section with semi-axes of 1.5 μm in the substrate plane and 2.75μm perpendicular to the wire long axis (Fig S-1a,b) and measured their hysteresis loops using MOKE. The loops are found to exhibit a similar coercivity to our smaller nanowires as expected, due to the flat variation of $H_c$ in nanowires with diameter above 200nm [27]. However, additional transitions above the main switching field are not observed, lending credibility to our proposed explanation of the magnetometry in smaller structures.

Additional evidence demonstrating the role of the 3D nanostructured vertex in determining the magnetic properties of the smaller structures is seen within spin-SEM images of the vertex taken at remanence after the application of a 250mT field perpendicular to the substrate plane (Fig 3i,j,k). The images show, that despite the application of field far above the measured coercive field in the perpendicular direction (27mT), a complex micro-magnetic configuration consisting of several domain walls are found to remain pinned at the vertex, demonstrating that the local potential landscape has been shaped by the 3D nanostructured geometry.

## 3. Conclusions

To conclude, we have used two-photon lithography and electrodeposition in order to fabricate vertical magnetic nanowires, angled magnetic nanowires and complex 3D tetrapod nanostructures. We have demonstrated that the domain structure within 3D magnetic nanostructures of complex geometry can be imaged using spin-SEM and standard surface sensitive magnetometry techniques such as MOKE can measure the magnetisation reversal upon the 3D nanostructure surface. Shape anisotropy alone cannot explain the shape of the hysteresis loops. It must be concluded that the magnetocrystalline anisotropy in these polycrystalline Co structures and in particular the spread of easy axes determine domain patterns as well as magnetisation switching. Both MOKE and spin-SEM measurements suggest the 3D nanostructured vertex plays an important role in magnetisation process. Further improvements in feature size allowing the fabrication of single domain structures may be realised using photoresists optimised for deep UV exposure, shorter wavelength lasers [29] and advanced stimulated emission-depletion exposure.

**Methods**

*Fabrication:* A positive resist (AZ9260) was spun onto a glass/ITO (700 nm) substrate yielding a thickness of approximately 6 mm. A two-photon lithography system consisting of a 780 nm, 120 mW laser with pulse width 120 fs and rep rate 80 MHz was used to write structures within the resist. Samples were made where the laser power was varied between 3 – 10.5 mW, scan speed was varied between 5 – 20 mm s-1 and development time varied between 15-120 minutes. After development, electrodeposition was used to fill the channels with Co. A standard Watts bath (600 ml) was used consisting of cobalt sulphate (90 g), cobalt chloride (27 g), boric acid (14 g) and sodium laurell sulphate (1 g). A simple two-electrode implementation was used with a Co anode and operating at constant current of 1 mA. After electrodeposition, the resist was lifted off for 24 hours in acetone, after which samples were subject to 1-hour oxygen plasma to remove residual resist.

*Magnetometry*: A 150 mW, 650 nm laser was attenuated to a power of approximately 50 mW, expanded to a diameter of 1 cm, and passed through a Glan-Taylor polarizer to place the beam into s-polarisation. The beam was then focused onto the sample using an achromatic doublet (f=30 cm), leading to a spot size of approximately 50 μm². The reflected beam was collected using an achromatic doublet (f=10 cm) and passed through a second Glan-Taylor polarizer, from which the transmitted and reflected beams were directed into two amplified Si photodetectors, yielding the Kerr and reference signals respectively. A variable neutral density filter was used to ensure the reference and Kerr signals were of similar value. Subtraction of the reference from the Kerr signal compensates for any change in laser intensity drift and also removes any small transverse Kerr effect from the signal.

*Magnetic Imaging*: We employ spin-SEM (also known as SEMPA) [30] to investigate the domain patterns in these 3D structures with high spatial resolution. This technique is an off-spring of standard scanning electron microscopy which is equipped with a spin analyser. A focused beam of electrons (energy 8 keV) scans along the surface, thereby exciting a wealth of low energetic secondary electrons through electron-electron scattering. These electrons (energy 0-20 eV) are ejected into vacuum and subsequently spin-analysed. In a ferromagnetic material, the electron spin direction is a direct measure of the magnetisation direction in the top 1 nm of the ferromagnet. Because of its high surface sensitivity, the experiment is done in ultrahigh vacuum (1x10$^{-10}$ mbar), including preparation of a clean surface by removal of nonmagnetic contaminants by mild ion bombardment (Kr+ ions, 2000 eV energy). Our setup is capable of detecting two magnetisation components simultaneously, e.g. x- and y-component.

*Atomic force microscopy*: This was carried out using a Bruker Dimension 3000 microscope and commercial atomic force microscopy tips.

# Acknowledgements

SL gratefully acknowledges funding from EPSRC (EP/L006669/1, EP/P510750/1, EP/P511122/1). JGR and Y-LDH acknowledge financial support from the ERC advanced grant 247462 QUOWSS and EPSRC grant EP/M009033/1. The research leading to these results has received funding from the European Union Seventh Framework Programme [FP7-People-2012-ITN] under grant agreement 316657 (SpinIcur). M. Hunt and G.I. Williams contributed equally to this work. Information on the data that underpins the research reported here, including how to access them, can be found in the Cardiff University data catalogue at http://doi.org/10.17035/d.2017.0031438135.

# Electronic Supplementary Material

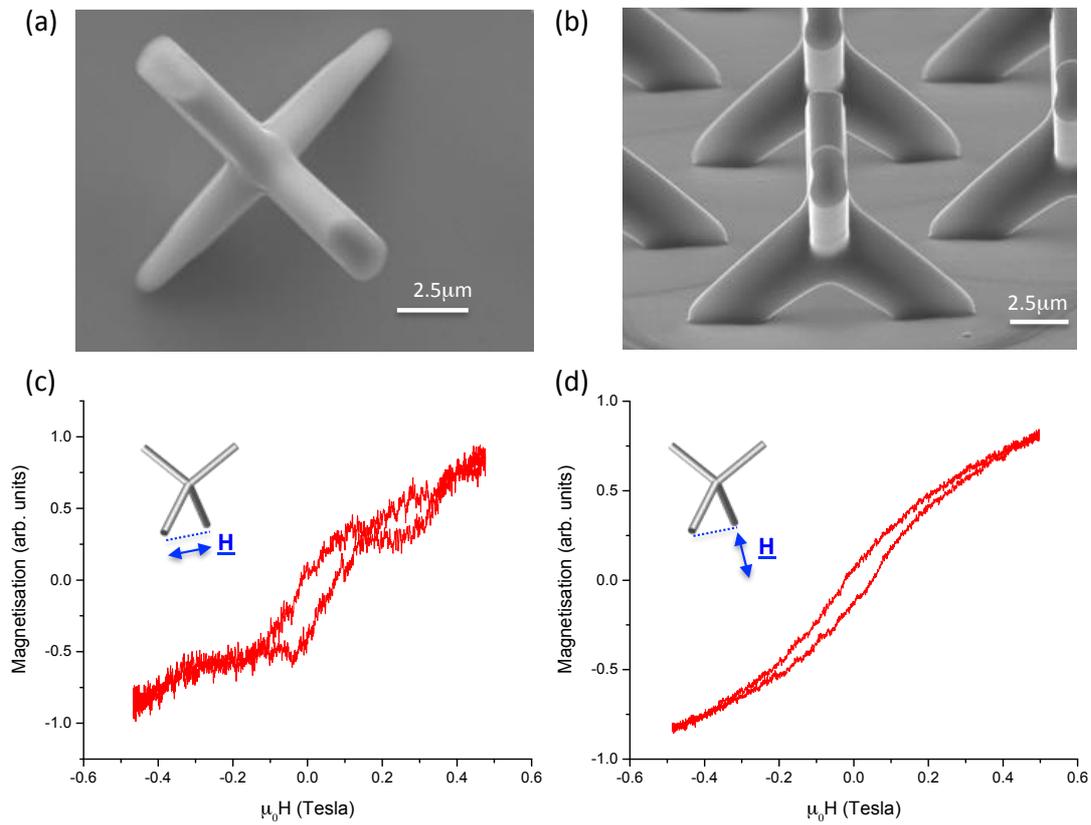

**Figure S-1. Tetrapod structures of larger feature size.** (a) Scanning electron microscopy image of larger tetrapod structures (Top view) (b) Scanning electron microscopy image of a larger tetrapod taken after 45° out-of-plane substrate rotation. Longitudinal MOKE loop of large tetrapod array with (c) field applied along projection of lower wires and (d) tetrapod array with field applied along projection of upper wires.